\documentstyle[prc,preprint,tighten,aps]{revtex} 
\begin{document}
\draft
\preprint{\vbox{\noindent 
 \hfill LA-UR-96-4914\\
 \null\hfill nucl-th/9701006 }}
\title{Nature of the first excited state of $\bbox{^4}$He} 
\author{Attila Cs\'ot\'o\thanks{csoto@qmc.lanl.gov} and 
G.~M. Hale\thanks{gmh@t2.lanl.gov}}
\address{Theoretical Division, Los Alamos National 
Laboratory, Los Alamos, New Mexico 87545} 
\date{January 2, 1997}

\maketitle

\begin{abstract}
\noindent
We study the first excited state of $^4$He in a microscopic 
\{$^3{\rm H}+p$,$^3{\rm He}+n$\} cluster model, including 
$^3$H and $^3$He distortions. The phenomenological $^1S_0$ 
$^3{\rm H}+p$ scattering phase shift is well reproduced. We 
localize a complex pole of the $S$-matrix between the 
$^3{\rm H}+p$ and $^3{\rm He}+n$ thresholds. The 
corresponding resonance parameters are $E_r=93$ keV 
position relative to $^3{\rm H}+p$, and $\Gamma=390$ keV 
width. A pole search is also performed in an extended 
$R$-matrix method, and a resonance is found with parameters 
$E_r=114$ keV and $\Gamma=392$ keV. The $R$-matrix approach 
gives several additional poles, some of which may be 
connected with an enhanced threshold effect.
\end{abstract}
\narrowtext

\section{Introduction}

$^4$He is the lightest nucleus with a well established 
spectrum of excited states. Thus it is an excellent testing 
ground for nuclear many-body models. Exact four-body 
calculations with realistic nucleon-nucleon ($N-N$) 
interactions have been performed for the $0^+$ ground state 
of $^4$He by using variational- \cite{VMC} and Green's 
function Monte Carlo \cite{GFMC} techniques, by solving the 
Yakubovsky equations \cite{Kamada}, and by using the 
correlated hyperspherical harmonic expansion method 
\cite{Viviani}. A few excited $^4$He states have also been 
studied by a variational Monte Carlo method \cite{Joe}. 
However, these latter calculations were performed without 
the Coulomb interaction, which fact made the theoretical, 
as well as the Coulomb-corrected experimental, $0^+_2$ 
state particle stable. So, the resonant nature of this 
state could not be studied. 

The nuclear shell-model offers another fundamental, and in 
principle, exact approach to calculate nuclear spectra. 
Early shell-model calculations, restricted to $(0+1)\hbar
\omega$ excitations, failed to provide a satisfactory 
description of both the ground state properties and the 
excitation spectrum of $^4$He \cite{01}. It was realized 
that higher $\hbar\omega$ excitations play important role 
in $^4$He, especially in the excited states 
\cite{Bevelacqua}. The shell-model calculations were 
substantially improved by using a $10\hbar\omega$ model 
space with various $N-N$ interactions \cite{Ceuleneer}. 
Those authors extensively studied the question whether the 
ground- ($0^+_1$) and first excited ($0^+_2$) states 
can be described simultaneously in a consistent way. It 
turned out that using a harmonic-oscillator size parameter 
that is optimal for the ground state in the calculations 
for the $0^+_2$ excited state made the excitation energy 
either several MeV too large or too small depending on the 
$N-N$ interaction. The usual resonance prescription puts 
this $0^+_2$ first excited state between the $^3{\rm H}+p$ 
and $^3{\rm He}+n$ thresholds at $E_r=395$ keV (all 
energies are given in this paper in the center-of-mass 
frame, relative to the $^3{\rm H}+p$ threshold) 
\cite{Tilley}; thus, delicate effects of few-body dynamics 
are expected to play an important role. The $0^+_2$ state 
is generally viewed as a one-particle--one-hole 
``breathing'' excitation of the ground state. In Ref.\ 
\cite{Ceuleneer} it was found that the charge radius of 
the $0^+_2$ state is significantly larger than that of the 
$0^+_1$ ground state, while the $D$-state probabilities 
are similar. The authors interpreted these findings as 
support for the breathing-mode interpretation.

Recently a series of calculations have been performed for 
$^4$He and for other light nuclei in large, no-core shell 
models, using interactions derived from realistic $N-N$ 
forces \cite{Zheng1}. The model reproduced the experimental 
$^4$He spectrum rather well, except for the $0^+_2$ state 
\cite{Zheng1}. This state was not the first excited state 
in that model as its excitation energy exceeded the 
experimental one by more than 10 MeV in a $4\hbar\omega$ 
shell-model space. In larger model spaces the $0^+_2$ state 
gradually moved toward lower energies, and in a 
$8\hbar\omega$ calculation it was found to be the second 
excited state, some 1.5 MeV higher in energy than 
experimentally \cite{Zheng2}. Its position relative to the 
3+1 thresholds could not be determined because the 
starting-energy dependence of the model made comparisons 
between the energies of different nuclei rather ambiguous. 
Recently the starting-energy dependence was removed from 
that model, thus allowing excited states to be referenced 
to breakup thresholds \cite{Navratil}. The $0^+_2$ state 
was found to be the second excited state situated above 
the $^3{\rm He}+n$ threshold of that model \cite{Navratil}. 
Thus, its shell-model description still needs improvements. 

We would like to emphasize here that shell-model 
calculations use unphysical boundary conditions for 
scattering wave functions \cite{comment}. It means that by 
increasing the shell-model space, the energy of an unbound 
state converges not to the resonance energy, but to the 
lowest two-body threshold \cite{comment}, or to 
$^3{\rm H}+p$ in the present case. In such a case, the 
shell-model wave function describes a situation where 
three nucleons stay close to each other forming a triton, 
while the fourth nucleon (proton) is far away. Thus, the 
shell-model wave function mimics a scattering wave function 
that has a node at the spatial range of the last 
shell-model basis function. If the phase shift in the 
two-body channel shows a sharp increase at a given energy, 
which is a sign of a resonance, then the shell-model 
energy is almost stable with respect to large variations 
of the spatial range of the basis, $i.e.$, the size of the 
shell-model space. This means that the shell-model energy 
has as a function of the size of the model space a plateau, 
from which, in principle, the position and width of the 
resonance can be extracted \cite{Hazi}. 

The origin of the great difficulties of the shell-model to 
reproduce the $0^+_2$ state at the correct excitation 
energy, between the $^3{\rm H}+p$ and $^3{\rm He}+n$ 
thresholds, is obvious. The closeness of this state to 
those two-body channels means that the most relevant 
degrees of freedom are the $^3{\rm H}+p$ and $^3{\rm He}+n$ 
relative motions. Thus, configurations that describe the 
$^3{\rm H}+p$ and $^3{\rm He}+n$ clustering have large 
weight in the wave function. It is known that wave 
functions which explicitly contain two-body (or three-body) 
clustering, correspond to shell-model states with very high 
$\hbar\omega$ excitations \cite{Suzuki}. The shell model 
treats all degrees of freedom equally, so it requires very 
large model spaces to correctly reproduce the two-cluster 
correlations.

In the present work we study the $0^+_2$ state of $^4$He 
in a microscopic cluster model. Contrary to the 
shell-model, this approach emphasizes the two-cluster 
correlations by building up the wave function from 
configurations with two-body dynamical degrees of freedom. 
Recently the spectrum of $^4$He has been extensively 
studied in a cluster model \cite{Hofmann}. We do not repeat 
here all those calculations. Our prime target is the 
$0^+_2$ state, and we also study the $0^+_1$ ground state. 
The nature of the $0^+_2$ state is not well understood. 
For example, in \cite{Tilley} it was speculated that this 
state might originate from an $S$-matrix pole far away from 
the physical region, and several MeV higher in energy than 
its 395 keV experimental excitation energy would suggest. 
In the present work we study the problem at complex 
energies, and try to reveal the pole structure of the 
$S$-matrix. The same method has recently been used to 
study the $3/2^-$ and $1/2^-$ low-lying states of $^5$He 
and $^5$Li \cite{He5}. Some further details can be found 
there.

\section{Resonating group model (RGM)}

We use a microscopic two-cluster Resonating Group Model 
(RGM) approach to $^4$He. The trial function of the 
four-body system is
\begin{eqnarray}
\Psi&=&\sum_{i=1}^{N_t}\sum_{S,L}
{\cal A}\Bigg \{\bigg [ \Big [(\Phi ^{t_i}\Phi^p) \Big ]_S 
\chi_L^{t_ip}(\bbox{\rho}_{tp})\bigg ]_{JM} \Bigg \} + 
\sum_{i=1}^{N_h}\sum_{S,L}
{\cal A}\Bigg \{\bigg [ \Big [(\Phi ^{h_i}\Phi^n) \Big ]_S 
\chi_L^{h_in}(\bbox{\rho}_{hn})\bigg ]_{JM} \Bigg \} \cr 
& &+
\sum_{i,j=1}^{N_d}\sum_{S,L}
{\cal A}\Bigg \{\bigg [ \Big [(\Phi ^{d_i}\Phi^{d_j}) 
\Big ]_S \chi_L^{d_id_j}(\bbox{\rho}_{dd})\bigg ]_{JM} 
\Bigg \}, 
\label{wf}
\end{eqnarray}
where ${\cal A}$ is the intercluster antisymmetrizer, the 
$\bbox{\rho}$ vectors are the various intercluster Jacobi 
coordinates, $L$ and $S$ is the total angular momentum and 
spin, respectively, and [...] denotes angular momentum 
coupling. While $\Phi^p$ and $\Phi^n$ is a neutron and 
proton spin-isospin eigenstate, respectively, the 
antisymmetrized ground state ($i=1$) and continuum excited 
distortion states ($i>1$) of the $t$, $h$, and $d$ clusters 
($t=\:$$^3$H, $h=\:$$^3$He, and $d=\:$$^2$H) are 
represented by the wave functions
\begin{eqnarray}
\Phi ^{t_i}&=&
\sum_{j=1}^{N_t}A^t_{ij}\phi ^t _{\beta _j},\ \ 
i=1,2,...,N_t, \cr
\Phi ^{h_i}&=&
\sum_{j=1}^{N_h}A^h_{ij}\phi ^h _{\beta _j},\ \ 
i=1,2,...,N_h, \cr
\Phi ^{d_i}&=&
\sum_{j=1}^{N_d}A^d_{ij}\phi ^d _{\beta _j},\ \ 
i=1,2,...,N_d.
\label{states}
\end{eqnarray}
Here $\phi ^t _{\beta _j}$, $\phi ^h _{\beta _j}$, and 
$\phi ^d_{\beta _j}$ are translationally invariant $0s$ 
harmonic-oscillator shell-model wave functions of $t$, $h$, 
and $d$, respectively, with size parameter $\beta _j$, and 
the $A_{ij}$ parameters are to be determined by minimizing 
the energies of the free clusters \cite{Tang}. Our choice 
of $L$, $S$, $N_t$, $N_h$, and $N_d$ will be discussed 
later. Putting (\ref{wf}) into the four-nucleon 
Schr\"odinger equation which contains a two-nucleon strong 
and Coulomb interaction, we get an equation for the 
intercluster relative motion functions $\chi$. For bound 
states, these relative motion functions are expanded in 
terms of square-integrable tempered Gaussian functions 
\cite{Kamimura}, and the expansion coefficients are 
determined from a variational method. For scattering 
states, we employ a Kohn-Hulth\'en variational method for 
the $S$-matrix, which uses square-integrable basis 
functions matched with the correct scattering asymptotics 
\cite{Kamimura}.

\section{Results and discussion}
\subsection{RGM}

For the $N-N$ force we use the Minnesota (MN) effective 
interaction \cite{MN} together with the tensor force of 
\cite{Heiss}. The MN interaction reproduces the deuteron 
binding energy in a $^3S_1$ model space without the 
$D$-state, which means that the interaction is too strong 
in the $^3S_1$ partial wave \cite{beta}. For the ground- 
and first excited $J^\pi=0^+$ states we use $L=S=0$ in the 
$^3{\rm H}+p$ and $^3{\rm He}+n$ configurations, and 
$L=S=0$ and $L=S=2$ in the $d+d$ configurations (the 
deuteron spin is 1). The $d+d$ configurations contain 
$^3S_1$ $N-N$ states, so we can expect that a model space 
which contains these configurations leads to unphysical 
overbinding. We use four different model spaces: (i) 
$N_t=N_h=1$ and there is no $d+d$ component; (ii) 
$N_t=N_h=3$ and there is no $d+d$ component; (iii) 
$N_t=N_h=3$, $N_d=1$ and only the $L=S=0$ state is present 
in the $d+d$ channel; (iv) $N_t=N_h=3$, $N_d=1$ and both 
the $L=S=0$ and $L=S=2$ states are present in the $d+d$ 
channel. The $E_{tp}-E_{hn}$ threshold energy difference is 
0.744 MeV and 0.770 MeV for the $N_t=N_h=1$ and 
$N_t=N_h=3$ model spaces, respectively. It is to be 
compared to the 0.763 MeV experimental value. Our $d+d$ 
threshold is 6.8 MeV above the $^3{\rm H}+p$ one in the 
$N_t=N_h=3$ model, while the experimental value is 4.0 MeV. 

Due to the $^3S_1$ overbinding problem, model spaces (iii) 
and (iv) are rather unphysical. We use them only for test 
purposes; that is why we do not allow for distortions in 
the deuteron clusters in Eq.\ (\ref{states}). We note, 
that Ref.\ \cite{Hofmann} used an $N-N$ interaction which 
was free from the above defect, so the model worked in a 
full $\{$$^3{\rm He}+p$,$^3{\rm He}+n,d+d\}$ space without 
any problem. Our main purpose is to localize the $0^+_2$ 
state. Test calculations for other systems, for instance 
for the $3/2^+$ state of $^5$He \cite{dt,phd}, show that 
in order to reproduce experimental resonance parameters, 
one really needs to reproduce only the relevant 
experimental phase shifts. This is true even if the 
description of the free clusters is highly unphysical, 
$e.g.$, if they are unbound \cite{phd}.

The most relevant phase shift in the present problem is 
that for $^1S_0$ $^3{\rm H}+p$. In Fig.\ 1 we show this 
phase shift coming from the various model spaces. We note 
that both the phase shifts and the binding energies are 
almost totally insensitive to the mixing parameter $u$ of 
the $N-N$ interaction. We use $u=0.98$ and the 
corresponding variationally stabilized oscillator size 
parameters for the clusters. One can see in Fig.\ 1 that 
the effect of the $^3$H and $^3$He distortions is 
significant, and that the model spaces which contain $d+d$ 
components show the overbinding effect, as expected. Thus, 
our best model is model (ii) with $N_t=N_h=3$ and without 
the $d+d$ configurations.

First we perform calculations for $J^\pi=0^+$ states by 
applying bound state asymptotics in (\ref{wf}). This wave 
function satisfies the correct physical asymptotics only 
for states which are below all the breakup thresholds, 
$i.e.$, below the $^3{\rm H}+p$ threshold in the present 
problem. For states above this threshold, this is a bound 
state approximation, like in the shell-model. In Table I we 
show the positions of the two lowest $0^+$ states in the 
various model spaces together with the amounts of 
clustering of the various configurations. This latter 
quantity gives the probability that the wave function is 
entirely in the given subspace \cite{amcl,he6}. Thus it is 
a useful measure of the relative importance of 
non-orthogonal channels. 

The $0^+_1$ ground state is slightly overbound compared to 
the $E_r=-19.815$ MeV experimental value \cite{Tilley}. 
Each model space predicts the $0^+_2$ state between the 
$^3{\rm H}+p$ and $^3{\rm He}+n$ thresholds. One can see 
from the amounts of clustering of the various 
configurations, that in the ground state the $^3{\rm H}+p$ 
and $^3{\rm He}+n$ clusterizations are equally important, 
while $0^+_2$ is predominantly a $^3{\rm H}+p$ state. The 
point nucleon rms radius of the ground state is around 1.6 
fm in our model, only slightly larger than the 1.48 fm 
experimental value. However, the radius corresponding to 
the $0^+_2$ state is huge, being around 40 fm. This is an 
unphysical value, which shows that the bound state 
approximation to a state which is above breakup thresholds 
might not make much sense \cite{comment}. So, the 
conclusions of Ref.\ \cite{Ceuleneer} concerning the 
breathing mode are questionable.

For a reliable localization of a state above breakup 
thresholds, the correct scattering asymptotics in the 
various channels must be imposed. Then one can search for 
resonant states either by studying the phase shifts, or by 
exploring the pole structure of the scattering matrix. In 
order to avoid any ambiguity in the recognition of a 
resonance in the phase shift, we choose here the latter 
method. We solve the Schr\"odinger equation for the 
relative motion functions $\chi$ in Eq.\ (\ref{wf}) at 
complex energies with the following boundary conditions 
for $\rho\rightarrow \infty$ 
\begin{equation}
\chi^{ab}_L(\varepsilon,\rho_{ab})\rightarrow H_L^-(k
\rho_{ab})-S_L(\varepsilon) H_L^+(k\rho_{ab}). 
\end{equation}
Here $\varepsilon$ and $k$ are the {\em complex} energies 
and wave numbers of the relative motions between clusters 
$a$ and $b$, and $H^-$ and $H^+$ are the incoming and 
outgoing Coulomb functions, respectively. We search for 
the poles of $S$ by extending the coupled channel 
scattering approach of Ref.\ \cite{Kamimura} to complex 
energies \cite{dt,He5}. The complex Coulomb functions are 
calculated by using Ref.\ \cite{Thompson}. The resulting 
complex energies $\varepsilon$ of the poles are connected 
to the resonance parameters via
\begin{equation}
\varepsilon =E_r-i\Gamma/2,
\end{equation}
where $E_r$ is the position of the resonance, and $\Gamma$ 
is its width.

In the case of an $N$-channel scattering problem, the 
complex channel wave numbers $k_1,k_2,...,k_N$, which 
determine the character of a state (bound state, scattering 
state, resonance) can be mapped by a one-to-one mapping to 
the $2^N$-sheeted Riemann surface of the complex channel 
energies $\varepsilon_1,\varepsilon_2,...,\varepsilon_N$ 
\cite{Eden}. The sheets of this surface can be labeled by 
an $N$-term sign string given by the signs of the imaginary 
parts of the channel wave numbers 
$[{\rm sgn}({\rm Im}~k_1),{\rm sgn}({\rm Im}~k_2),...,{ 
\rm sgn}({\rm Im}~k_N)]$. It has been shown 
\cite{Eden,Pearce} that in the case of Hermitian 
potentials, a complex pole of the $S$-matrix that would 
appear in one of the $N$ channels in a single-channel 
problem, gives rise to $2^{N-1}$ poles on different Riemann 
sheets in the $N$- channel problem. The proof of this 
statement is based on the fact that in the zero coupling 
limit, when the only coupling is the energy conservation, 
the $N\times N$ Fredholm determinant of an $N$-channel 
scattering problem reduces to the product of $N$ 
one-channel Fredholm determinants. However, the situation 
is different if there are non-orthogonal channels, like 
the $(L,S)=(0,0)$ ones in the present case. Such channels 
are inherently coupled, and the zero coupling limit cannot 
be taken. In such cases one does not know the number and 
location of the poles, so one has to search all energy 
sheets.

Following \cite{Eden}, the poles lying on the sheet closest
to the physical sheet ($[++\cdots\,+]$) at a given energy, 
are called conventional poles, while the others are called 
shadow poles. Usually only the conventional poles have 
observable effects, causing the appearance of conventional 
resonances. However, there are exceptions where the effects 
of shadow poles are non-negligible or even dominant. We 
mention here the examples of the $^3$H($d,n)^4$He reaction 
\cite{Hale,dt} and the structure of $^8$Be \cite{Be8}. 
Shadow poles play an important role also in atomic physics, 
in laser ionization processes, because of the large number 
of channels and relatively low energies required for 
ionization \cite{Potvliege}. In Ref.\ \cite{Tilley} it was 
speculated that the $0^+_2$ state of $^4$He might come from 
a shadow pole, which fact could partly explain the 
difficulties encountered by the shell-model in reproducing 
this state at the correct energy.

In order to explore this possibility, we searched all 
energy sheets for poles. In model spaces that include 
$^3$H and $^3$He distortions, the continuum excited 
distortion states represent high-lying channels, $e.g.$, 
in model space (ii) we have six channels: 
$\{t_1+p,t_2+p,t_3+p,h_1+n,h_2+n,h_3+n\}$. The dominant 
Riemann sheets are those where the distortion channels all 
have bound state character ($+$), rather than antibound 
state character ($-$). In the case of model (ii) these 
dominant channels are $[+++++\:+]$, $[+++-+\:+]$, 
$[-++++\:+]$, and $[-++-+\:+]$. Numerical studies show 
\cite{phd} that if there is an $S$-matrix pole on one of 
these sheets, then the corresponding pole on a sheet, 
where the character of at least one distortion channel is 
``$-$'', is situated almost exactly at the same complex 
energy position as the original pole. Since these latter 
sheets are much farther from the physical region than the 
four dominant ones, their poles have negligible observable 
effect. That is why we set the character of all distortion 
channels to ``$+$'' and give only the characters of those 
channels that contain the ground states of the clusters; 
$e.g.$, $[--]$ means $[-++-+\:+]$ in model space (ii).

In model space (i) we do not find any pole, while in the 
(ii), (iii), and (iv) model spaces we find one pole on the 
$[-+]$, $[-++]$, and $[-++]$ sheets, respectively at 
$(0.093- i0.195)$ MeV, $(0.085-i0.071)$ MeV, and 
$(0.053- i0.021)$ MeV complex energies, respectively. 
(Note that the thresholds of the two $d+d$ channels 
coincide, so the character of the fourth channel is always 
the same as that of the third one.) One can see the effect 
of the $^3S_1$ overbinding problem in the pole positions 
in the (iii) and (iv) models. In each model space the pole 
is on the Riemann sheet which is closest to the physical 
sheet, $i.e.$, it is a conventional pole. We do not find 
any other pole on any other sheet in the vicinity of the 
$^3{\rm H}+p$ and $^3{\rm He}+n$ channel thresholds.

To recap the results of the RGM calculations, our best 
model with $^3$H and $^3$He distortions predicts the 
$0^+_2$ state to be a conventional resonance at $E_r=93$ 
keV above the $^3{\rm H}+p$ threshold, with 390 keV width. 

\subsection{R-matrix}

The RGM results encouraged us to search again for this 
state as an $S$-matrix pole in the charge-independent 
$R$-matrix analysis of reactions in the $A=4$ system 
reported in Refs.\ \cite{Tilley} and \cite{Hofmann}. The 
state was visible at $E=395$ keV according to the usual 
resonance-parameter prescription, but did not appear to 
give a low-lying $S$-matrix pole using the ``extended" 
$R$-matrix prescription \cite{Hale,He5}. This prescription 
involves first fitting the available experimental data in 
terms of the conventional $R$-matrix parametrization at 
real energies on the physical sheet, then using this 
parametrization to continue the $S$-matrix onto other 
sheets of the Riemann energy surface in order to study its 
analytic structure \cite{Hale}, in very much the same way 
as discussed above. 

The channel configuration and the distribution by reaction 
of data included in the $A=4$ $R$-matrix analysis are 
summarized in Table II, taken from Ref.\ \cite{Hofmann}. 
In general, all types of cross-section and polarization 
measurements were used, but the ones that showed most 
clearly the $0^+$ resonance and its associated threshold 
effect were excitation functions of the $^3$H($p,p)^3$H 
differential elastic cross section 
\cite{Ba59,Ba64,Ja59,Ja63,En54}. Some of those measurements 
are shown compared with the $R$-matrix calculation in 
Fig.\ 2. The resonance peak occurs at about 350 keV, and 
the threshold step at $E=764$ keV is especially striking 
at this angle ($\theta_{c.m.}=120^\circ$). The resulting 
$^1S_0$ $^3{\rm H}+p$ phase shift, represented by the dots 
in Fig.\ 1, serves as the ``experimental" data to which 
the RGM results are compared.

We find an $S$-matrix pole on the $[-++]$ sheet at 
$(0.114-i0.196)$ MeV, corresponding to a conventional 
resonance at $E_r=114$ keV above the $^3{\rm H}+p$ 
threshold, with $\Gamma=392$ keV width, in good agreement 
with the parameters obtained from the RGM. At the time of 
the $^4$He level compilation reported in \cite{Tilley}, 
this pole had not been found, leading to the speculation 
that the resonance might actually be associated with 
higher-lying shadow poles in the $0^+$ state. These shadow 
poles occur at energies between about 3.0 and 3.6 MeV, with 
widths in the range $6-8$ MeV, on the Riemann sheets 
$[-++]$, $[-+- ]$, $[+--]$, and $[+-+]$. In addition, 
there is another resonance at $\varepsilon=(7.68-i3.57)$ 
MeV on the $[---]$ sheet, with an associated shadow pole 
at $\varepsilon=(8.43-i3.43)$ MeV on the $[--+]$ sheet.

It is interesting to note that all of the $S$-matrix 
structure described above comes predominantly from the 
same $T=0$ level in the $R$-matrix, located approximately 
6 MeV above the $^3{\rm H}+p$ threshold.  The position of 
this level depends on the boundary conditions, which are 
taken to be the shift functions in the various 
channels\footnote{The boundary condition used in the 3+1 
channels is actually the average of the $^3{\rm H}+p$ and 
$^3{\rm He}+n$ shifts, in order to preserve the 
charge-independent model.} evaluated at the ground-state 
energy of $^4$He, so that the lowest $0^+$ $R$-matrix level 
coincides identically with the $^4$He ground state. One 
can then imagine the 6-MeV level to be associated with a 
small-basis shell-model wave function, since these states,
like the internal $R$-matrix eigenfunctions, are expected 
to represent the true wave function of the scattering 
system only in a limited region of space.  The point is 
that, when such an expansion is matched to the correct 
asymptotic scattering solution, it produces a low-lying 
$S$-matrix pole in the correct position for the resonance 
associated with the first excited state of $^4$He, even 
though the energy eigenvalue of the structure state is far 
above the resonance energy.  Of course, as was discussed 
earlier, enlarging the shell-model basis would make the 
energy of the state decrease until, at some point, it 
would pass through the resonance energy on the way to 
attaining its minimum value (the $^3{\rm H}+p$ threshold 
energy).  However, the correct information about the 
resonance as an $S$-matrix pole may already be contained 
in the small-basis shell-model states.

As was noted earlier, the additional poles above the 
$^3{\rm He}+n$ threshold were not found in the RGM 
approach. Compared to the $R$-matrix model, our RGM 
approach is less realistic, mainly because the description 
of the $d+d$ channels is rather schematic due to the 
$^3S_1$ force problem. On the other hand, the $R$-matrix 
approach embodies some aspects of channel orthogonality in 
the region outside the nuclear surface that might increase 
the likelihood of having multiple poles ($cf$.\ the 
discussion about the number of poles). The nature of these 
additional states in the $R$-matrix spectrum will require 
further investigation. However, they appear to be necessary 
in order to enhance the strong threshold step that is seen 
in the measured $^3{\rm H}+p$ cross-section excitation 
functions. 

Both calculations are in substantial agreement that the 
first excited state of $^4$He is $0^+_2$, a conventional 
resonance lying between the $^3{\rm H}+p$ and 
$^3{\rm He}+n$ thresholds, with $E_r\approx 100$ keV energy 
relative to $^3{\rm H}+p$, and $\Gamma\approx 400$ keV. 
Since the real-energy resonance parameters for this state, 
$E_r=395$ keV and $\Gamma=500$ keV \cite{Tilley}, were 
obtained from the same $R$-matrix parameters as used here, 
the differences come entirely from the relation of 
resonance parameters determined from real- and 
complex-energy scattering quantities, respectively, as 
discussed in Ref.\ \cite{He5}.

\section{Conclusion}

In summary, we have described the $0^+_1$ and $0^+_2$ 
states of $^4$He in a microscopic 
\{$^3{\rm H}+p$,$^3{\rm He}+n$\} RGM approach. The 
effective interaction did not allow us to fully include 
$d+d$ configurations into the model. We have found that 
$^3$H and $^3$He cluster distortions play important roles 
if one wants to reproduce the relevant $^1S_0$ 
$^3{\rm H}+p$ phase shift. Our best model, which 
satisfactorily reproduced this phase shift, put the ground 
state of $^4$He at --20.53 MeV relative to the 
$^3{\rm H}+p$ threshold. We have searched for $S$-matrix 
poles at complex energies using the same model space, and 
found one at $(0.093-i0.195)$ MeV energy, relative to the 
$^3{\rm H}+p$ threshold. Our model predicts that this 
$0^+_2$ state is the first excited state of $^4$He, and is 
a conventional resonance at $E_r=93$ keV with $\Gamma=390$ 
keV width. While in the ground state both the $^3{\rm H}+p$ 
and the $^3{\rm He}+n$ configurations have roughly the 
same weight, the $0^+_2$ state is dominated by the 
$^3{\rm H}+p$ configuration. 

We have also localized the $0^+_2$ state in an extended 
$R$-matrix model. Its parameters $E_r=114$ keV and 
$\Gamma=392$ keV are in good agreement with the RGM 
parameters. The $R$-matrix model produces several 
additional $0^+$ poles. While the understanding  of these 
structures will require further theoretical investigation, 
their  role in producing a strong threshold effect is 
already clearly seen.

\acknowledgments

This work was performed under the auspices of the U.S.\ 
Department of Energy. Early stages of this work were done 
at the National Superconducting Cyclotron Laboratory at 
Michigan State University, supported by NSF Grants.\ 
PHY92-53505 and PHY94-03666. Support from OTKA Grant 
F019701 is also acknowledged.

\mediumtext
\begin{figure}
\caption{$^1S_0$ phase shifts for $^3{\rm H}+p$ scattering, 
coming from model spaces (i) -- dotted line, (ii) -- solid 
line, (iii) -- dash-dotted line and (iv) -- dashed line. 
Model spaces (i)--(iv) are defined in the text. The solid 
dots come from an $R$-matrix analysis of the experimental 
data \protect\cite{Hofmann}.}
\label{fig1}
\end{figure}

\mediumtext
\begin{figure}
\caption{Differential cross section for $^3$H$(p,p)^3$H 
elastic scattering at $\theta_{c.m.} \approx 120^\circ$. 
The solid curve is the $R$-matrix calculation, and the 
data are from Refs.\ \protect\cite{Ba59,Ba64} (solid 
circles), \protect\cite{Ja63} (open circles), and 
\protect\cite{En54} (solid triangles).}
\label{fig2}
\end{figure}

\widetext
\begin{table}
\caption{Energies (relative to $^3{\rm H}+p$) of the 
$0^+_1$ and $0^+_2$ states of $^4$He, and the amounts of 
clustering of the various cluster configurations in these 
states in model spaces (i)--(iv), defined in the text. The 
three numbers in parentheses are for the three $^3$H or 
$^3$He states, in the case of $N_t=3$ or $N_h=3$. In the 
$d+d$ channels the $(L,S)$ values are also given.} 
\begin{tabular}{lr@{}lllr@{}lll}
& \multicolumn{4}{c}{$0^+_1$} &
\multicolumn{4}{c}{$0^+_2$} \\
\cline{2-5}
\cline{6-9}
Model
& \multicolumn{2}{c}{$E$ (MeV)}
& \multicolumn{2}{c}{Amount of clustering} & 
\multicolumn{2}{c}{\ \ \ \ $E$ (MeV)}
& \multicolumn{2}{c}{Amount of clustering} \\ 
\tableline
(i) & --20.&83 & \ \ $^3{\rm H}+p$ & 97.5 & \ \ \ \ 0.&54 & \ \
$^3{\rm H}+p$ & 90.0 \\
&       & & \ \ $^3{\rm He}+n$ & 97.3 & & & \ \
$^3{\rm He}+n$ & 11.9 \\
(ii) & --20.&53 & \ \ $^3{\rm H}+p$ & (94.8,10.5,0.05) & 0.&34 
& \ \ $^3{\rm H}+p$ & (80.4,16.5,0.07) \\ & 
& & \ \ $^3{\rm He}+n$ & (94.5,11.5,0.05) & & & \ \ $^3{\rm 
He}+n$ & (27.8,19.7,0.08) \\ (iii)& -- 20.&66 & \ \ $^3{\rm 
H}+p$ & (94.6,10.5,0.04) & 0.&24 & \ \ $^3{\rm H}+p$ & 
(76.2,20.0,0.08) \\ & 
& & \ \ $^3{\rm He}+n$ & (94.2,11.5,0.05) & & & \ \ $^3{\rm 
He}+n$ & (32.6,23.3,0.09) \\ & & & \ \ $d+d$ $(0,0)$ & 59.5 & 
& &
\ \ $d+d$ $(0,0)$ & 27.2 \\
(iv) & --21.&63 & \ \ $^3{\rm H}+p$ & (93.1,10.7,0.04) &
0.&15 & \ \ $^3{\rm H}+p$ & (73.8,22.9,0.09) \\ & 
& & \ \ $^3{\rm He}+n$ & (92.7,11.7,0.04) & & & \ \ $^3{\rm 
He}+n$ & (36.1,26.2,0.09) \\ & & & \ \ $d+d$ $(0,0)$ & 58.3 & 
& &
\ \ $d+d$ $(0,0)$ & 30.4 \\
&       & & \ \ $d+d$ $(2,2)$ & 1.4 & & &
\ \ $d+d$ $(2,2)$ & 0.3 \\
\end{tabular}
\end{table}

\mediumtext
\begin{table}
\caption{Channel configuration (top) and data summary 
(bottom) for each reaction in the $^4$He system $R$-matrix 
analysis. The maximum orbital angular momentum allowed for 
each arrangement is given by $l_{max}$, while $a_c$ is the 
channel radius.}
\begin{tabular}{cccc}
Channel & \multicolumn{2}{c}{$l_{max}$} & $a_c$ (fm) \\ 
\hline
$^3{\rm H}+p$ & \multicolumn{2}{c}{3} & 4.9 \\ $^3{\rm 
He}+n$ & \multicolumn{2}{c}{3} & 4.9 \\ $d+d$ & 
\multicolumn{2}{c}{3} & 7.0 \\
\hline
Reaction        & Energy range (MeV) & \# Observable
types & \# Data points \\
\hline
$^3$H$(p,p)^3$H & $E_p=0-11$    &       ~3 & 1382\\
$^3$H$(p,n)^3$He + $^3$He$(n,p)^3$H & $E_p=0-11$ 
& ~5 & ~726\\
$^3$He$(n,n)^3$He& $E_n=0-10$ & ~2 & ~126\\
$^2$H$(d,p)^3$H & $E_d=0-10$    &       ~6 & 1382\\
$^2$H$(d,n)^3$He & $E_d=0-10$ & ~6 & ~700\\
$^2$H$(d,d)^2$H & $E_d=0-10$    &       ~6 & ~336\\
\hline
&       Totals: & 28 & 4652\\
\end{tabular}
\end{table}

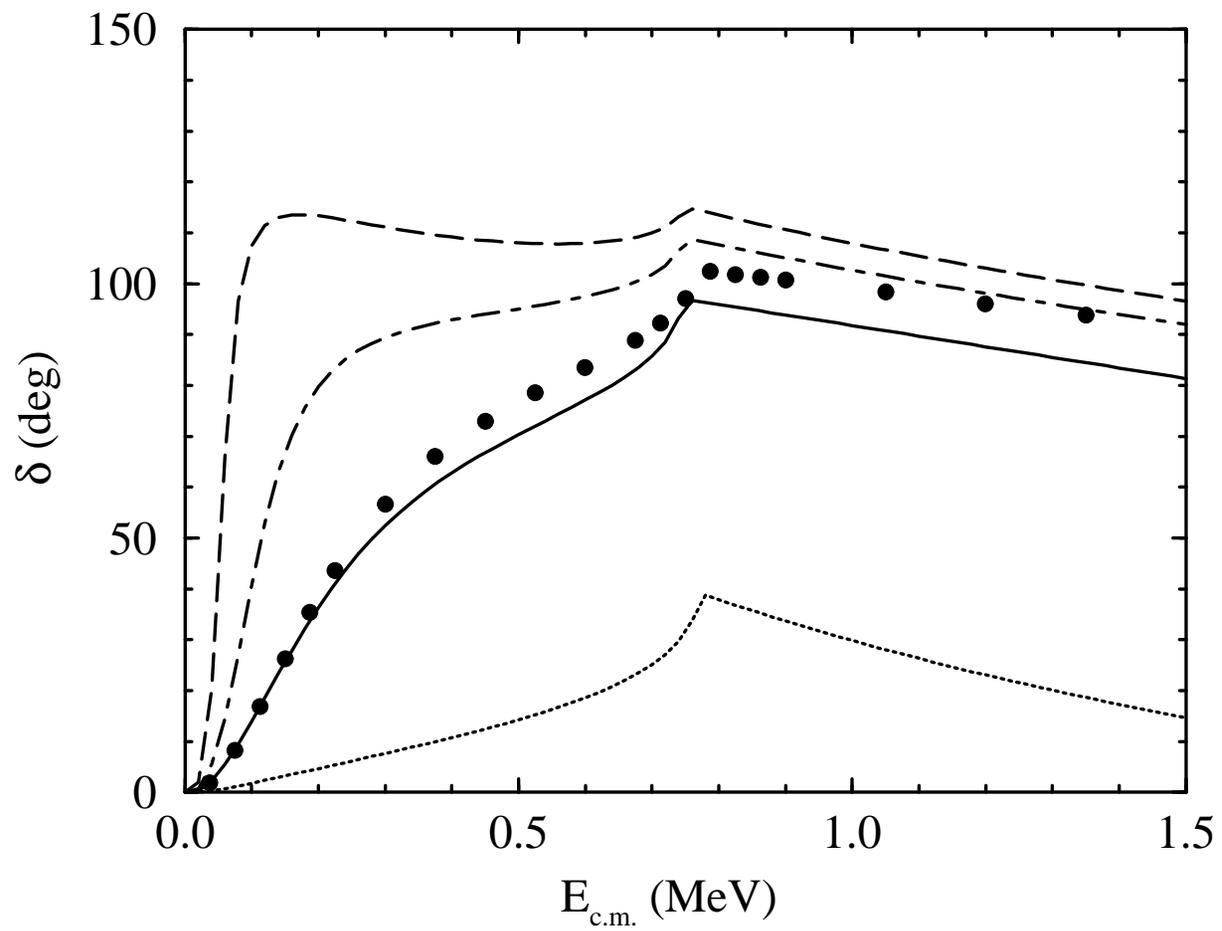
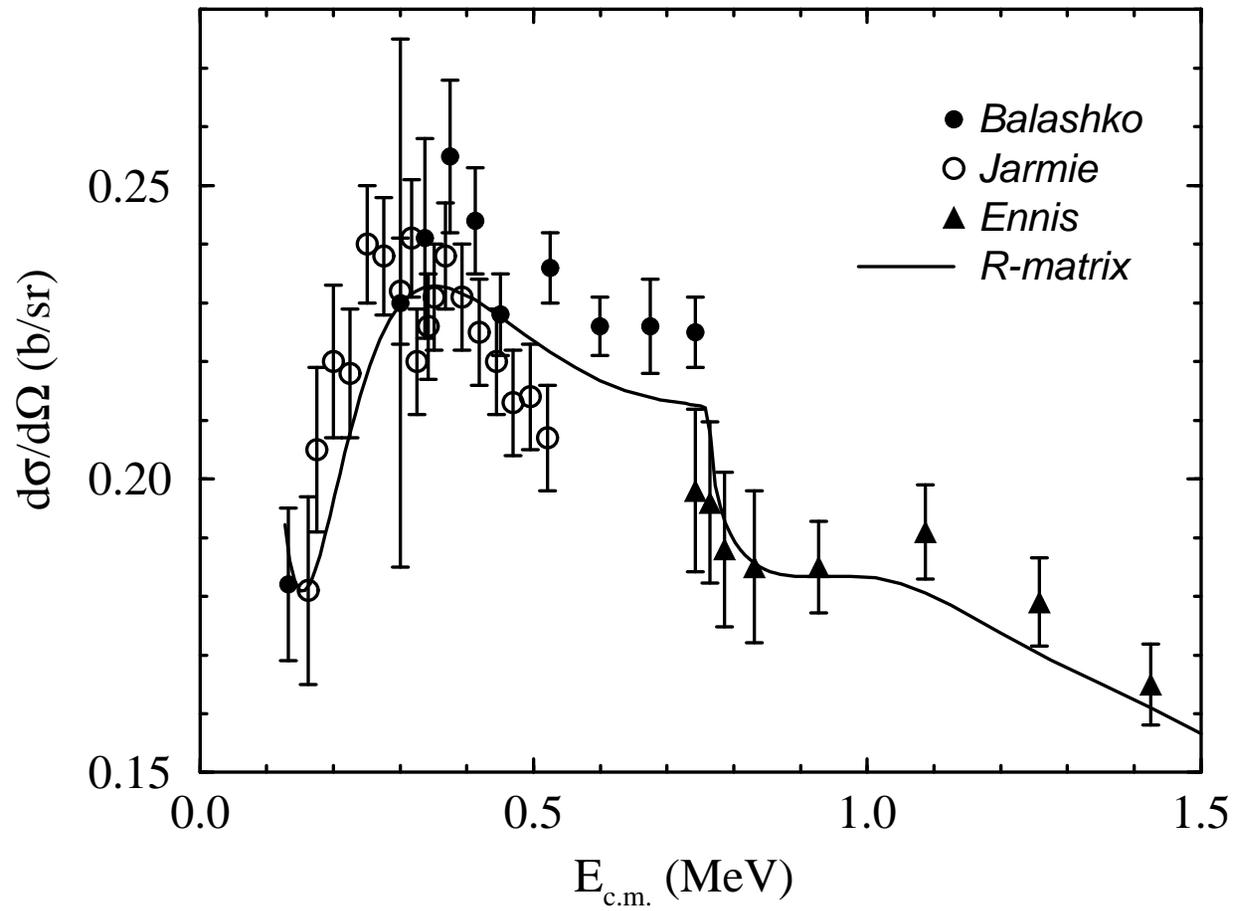
\end{document}